\documentclass[twocolumn,amsmath,amssymb]{revtex4}   

\usepackage{graphicx}
\usepackage{dcolumn}
\usepackage{bm}
\usepackage{amsfonts}

\usepackage[hyperindex,breaklinks]{hyperref}

\newcommand{\be}{\begin{equation}}
\newcommand{\ee}{\end{equation}}
\newcommand{\bea}{\begin{eqnarray}}
\newcommand{\eea}{\end{eqnarray}}

\newcommand{\cha}{\text{A}}
\newcommand{\chb}{\text{B}}
\newcommand{\chc}{\text{C}}

\newcommand{\nc}{{n_{\text{C}}}}
\newcommand{\nct}{$n_{\text{C}}$}
\newcommand{\ns}{{n_{\text{S}}}}
\newcommand{\nst}{$n_{\text{S}}$}

\newcommand{\cvect}{\mathbf{c}}
\newcommand{\fvect}{\mathbf{f}}
\newcommand{\yvect}{\mathbf{y}}
\newcommand{\ymat}{\mathbf{Y}}
\newcommand{\amat}{\mathbf{A}}
\newcommand{\mvect}{\mathbf{m}}
\newcommand{\zetavect}{\bm{\zeta}}

\newcommand{\kyyp}{k_{y \rightarrow y'}}
\newcommand{\kapyyp}{\kappa_{y \rightarrow y'}}

\newcommand{\pss}{P_{\text{SS}}}
\newcommand{\pois}{\mathcal{P}}

\newcommand{\mastw}{ \mathcal{W}}
\newcommand{\ymu}{y_{\mu}}

\begin{document}

\title{Equilibrium-like behavior in far-from-equilibrium chemical reaction networks}

\author{David K. Lubensky}
\affiliation{Department of Physics, University of Michigan, Ann Arbor MI  48109-1040}

\begin{abstract}
In an equilibrium chemical reaction mixture, the number of molecules present obeys a Poisson distribution.  We ask when the same is true of the steady state of a nonequilibrium reaction network and obtain an essentially complete answer.  In particular, we show that networks with certain topological features must have a Poisson distribution, whatever the reaction rates.  Such driven systems also obey an analog of the fluctuation-dissipation theorem.  Our results may be relevant to biological systems and to the larger question of how equilibrium concepts might apply to nonequilibrium systems.

\end{abstract}


\maketitle

Physicists have long sought to extend the  framework of equilibrium statistical mechanics to describe systems far from equilibrium.
Although this general goal is still far from attained, recent years have seen the discovery of several specific systems that, while strongly driven, nonetheless act very much as if they were in equilibrium.  Such unexpected thermal behavior has been observed, for example, in groups of fluidized spheres~\cite{durian}, in vibrated granular matter~\cite{olafsen}, and in slowly sheared collections of particles near the jamming transition~\cite{makse,liu-ohern}.  These cases are striking for the extent to which an equilibrium analogy holds; the dynamics of fluctuations and sometimes entire distributions of observables correspond to those expected in equilibrium.  Such observations serve as tantalizing hints that a simpler structure may be hiding under the apparent complexities of nonequilibrium statistical physics.  It is thus of considerable interest to understand why these examples act as they do and what the limits are of their likeness to thermal systems.  In the cases just cited, the equilibrium-like properties were found either experimentally or through large-scale simulations, and a full theoretical understanding of the observations has been slow to develop.  In particular, despite a few important contributions~\cite{DLevine}, it has proven difficult to say precisely when an equilibrium-like description should hold.  Here, we address this question for a new class of systems, chemical reaction networks, and show that in this case it can be answered in some detail.


The statistical physics of chemical systems has recently benefited from a resurgence of interest motivated by biological applications~\cite{noise-reviews}.   In some cells, regulatory molecules can be present in as few as one or two copies, on average.  In this regime, the intrinsic stochasticity of chemical reactions causes large fluctuations in molecule numbers, and it is natural to ask how cells can continue to function in the face of the resulting uncertainty.  A complete answer to this question requires knowing what determines the molecule number distribution in driven, reacting systems.  We take an important step in this direction by giving a full characterization of those reaction networks whose steady-state distribution (SSD) of molecule numbers is Poissonian, as it would be in an equilibrium system.  We show that a surprisingly large class of networks have this property.  Such networks can include completely irreversible reactions and an arbitrarily large number of chemical species; the network topology should, however, satisfy a sparseness condition that allows the construction of a generalized free energy landscape.  

In what follows, we first show that whether a mass-action network has a Poisson SSD (PSSD) is closely related to whether the mean-field kinetic equations for the same network have a unique fixed point.  This latter problem has a long history in the chemical engineering literature~\cite{feinberg,crnt-pedagog}, and by adapting some results from the deterministic case, we are able to give a mathematical description of the set of networks with a PSSD.  This set includes, for any choice of reaction rates, networks with a so-called deficiency zero topology.  We then show that the equilibrium analogy extends beyond the form of the SSD.  Networks with a PSSD satisfy a version of the fluctuation-dissipation theorem, and when two such systems are brought into contact, they exchange particles to equalize their values of a generalized chemical potential.  Analogous results hold for certain non-mass-action kinetic schemes.  A significant group of far-from-equilibrium networks thus shows a marked resemblance to equilibrium systems, and it is possible to gain considerable insight into when and why this occurs.

\begin{figure}

\includegraphics[width=8cm]{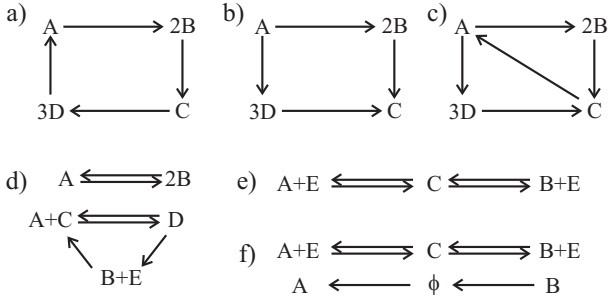}

\caption{ \label{fig1} Some chemical  reaction networks.  a)--c) have complexes $\{\text{A},\text{2B},\text{C},\text{3D}\}$ and $l = 1$ linkage classes.  d) has complexes $\{\text{A},\text{2B},\text{A+C},\text{D},\text{B+E}\}$ and $l = 2$.  e) has complexes $\{\text{A+E},\text{C},\text{B+E}\}$ and $l = 1$, f) complexes $\{ \text{A+E},\text{C},\text{B+E},\text{A},\phi,\text{B} \}$ and $l = 2$.  Networks a) and c)--e) are weakly reversible, and all except f) have deficiency $\delta = 0$.}

\end{figure}

We begin with some formal preliminaries~\cite{feinberg,crnt-pedagog}.  By a chemical reaction network, we mean a set of chemical \textit{species} (or \textit{molecules}) that can undergo a prescribed set of reactions (Fig.~\ref{fig1}).  For most of this paper, we consider a well-stirred reaction vessel, so that spatial degrees of freedom are irrelevant; later, we will argue that our results extend to models that explicitly include molecular diffusion.  We denote species by uppercase letters.
A \textit{complex} is the collection of species on one side of a reaction arrow together with their stoichiometric coefficients.  In the reaction $2\text{A} + \text{B} \rightarrow \text{C}$, the $\ns = 3$ species are $\{\cha,\chb,\chc\}$, and the $\nc = 2$ complexes are $\{2\cha + \chb,\chc\}$.   A network may include pseudo-reactions that, for example, absorb the concentration of a reactant that is in excess into the rate constant; thus, a reaction like $\phi \rightarrow A$, indicating that A molecules are created out of nothing, is possible.

Chemical reactions in dilute systems are most commonly modeled by deterministic rate equations (DRE).  In this description, the relevant variables are the concentrations $\cvect = (c_{\cha_1},c_{\cha_2},c_{\cha_3},\ldots)$ of the species $\{\cha_1,\cha_2,\cha_3,\ldots\}$, and these evolve according to $\dot{\cvect} = \fvect(\cvect)$.  If we assume mass action kinetics (as we do in most of this article), then $\fvect(\cvect) = \sum_{y,y'} k_{y \rightarrow y'} \cvect^{\yvect} (\yvect' - \yvect)$,
where $y$ and $y'$ denote reactant and product complexes, $k_{y \rightarrow y'}$ is the rate constant for the reaction $y \rightarrow y'$ (and is 0 if the reaction does not occur), and $\yvect$ is a vector of length \nst\ giving the stoichiometric coefficient $y_{\cha_j}$ of each species in complex $y$ [e.g. if $y = 2\cha_1 + 3\cha_2$, then $\yvect = (2,3,0,0,\ldots)$].  The expression $\cvect^{\yvect}$ is shorthand for $\prod_{j = 1}^{\ns} c_{\cha_j}^{y_{\cha_j}}$.  

This $\fvect(\cvect)$ can be rewritten in a form that will prove useful once we add stochastic effects:
\be
\fvect(\cvect) = \ymat \amat(k) \Psi(\cvect) \;. \label{eq:dre-matrices}
\ee
Here, $\ymat$ is an $\ns \times \nc$ matrix whose columns are the stoichiometric vectors $\yvect$ of the \nct\ complexes.  $\amat$ is an $\nc \times \nc$ matrix with entries (indexed by the names of the complexes)
\be
A_{y' \! ,y}(k) = \left\{
\begin{array}{cl}
k_{y \rightarrow y'}, & y \neq y' \\
-\sum_{y''} k_{y \rightarrow y''}, & y = y' \; . 
\end{array}
\right.
\ee
The argument $k$ indicates that $\amat(k)$ depends on the rate constants $k_{y\rightarrow y'}$ as well as on the network's topology.  Finally, $\Psi(\cvect)$ is a vector of length \nct\ with elements $\Psi_{y}(\cvect) = \cvect^\yvect$.

The DRE $\dot{\cvect} = \fvect(\cvect)$ gives a good account of the evolution of mean concentrations in macroscopic systems, but it hides the fact that individual reaction events occur stochastically.   A kinetic equation that incorporates this effect becomes important when fluctuations are large and is also essential to any statistical-mechanical treatment of reaction networks.  This equation is the chemical master equation (CME)~\cite{noise-reviews}; it specifies how the probability $P(\mvect, t)$ that there are $\mvect = (m_{\cha_1}, m_{\cha_2},m_{\cha_3},\ldots)$ molecules of the various species at time $t$ changes in time.  The CME takes the form
\bea
\frac{\partial P(\mvect, t)}{\partial  t} & = & \sum_{y,y'} \kapyyp \left[ \frac{ (\mvect + \yvect - \yvect')!}{ (\mvect - \yvect')!} P(\mvect + \yvect - \yvect') \right. \nonumber \\
&& \left. - \frac{\mvect!}{(\mvect - \yvect)!} P(\mvect) \right]  \nonumber \\
& \equiv & \mastw P(\mvect) \; , \label{eq:cme}
\eea
where, for a vector $\mathbf{v}$, $ \mathbf{v}! \equiv \prod v_j!$, with the product taken over all components $v_j$, and the factors $\mvect!/(\mvect - \yvect)!$ are the generalization to discrete molecule numbers of $\cvect^\yvect$ in the DRE.  The rate constants $\kapyyp$ differ from the constants $\kyyp$ in the DRE for the same network by a factor of $V^{-1 + \sum_j y_{\cha_j}}$, where $V$ is the system volume.

Because Eq.~\ref{eq:cme} is a master equation, it must have a SSD $\pss(\mvect)$ satisfying $\mastw \pss(\mvect) = 0$; with trivial exceptions, $\pss$ is unique~\cite{van-kampen}.  For a dilute chemical system in equilibrium with a heat bath, we know that $\pss$ must take the Poisson form
\be
\pss(\mvect) = \frac{1}{Z} \frac{ \zetavect^\mvect}{\mvect!} \equiv \pois(\mvect; \zetavect) \; . \label{eq:pssd}
\ee
The individual components $\zeta_{\cha_j}$ of $\zetavect$ include factors of the volume $V$ and of the thermal wavelength and may incorporate a partition sum over internal molecular degrees of freedom.  In general, a reaction network has conservation laws of the form $\yvect \cdot \mvect = \text{constant}$, and the constant $Z$ is chosen so that $\pss$ is normalized for given values of the conserved quantities.

We now ask whether any far-from-equilibrium reaction networks have a SSD of the Poisson form (\ref{eq:pssd}).  We can answer this question simply by substituting a Poisson distribution into the equation $\mastw \pss(\mvect) = 0$.  One finds that
\bea
\mastw \pois(\mvect;\zetavect) \! & \propto & \! \sum_{y,y'} \kapyyp \left[ \frac{\zetavect^{\mvect + \yvect - \yvect'}}{(\mvect - \yvect')!} - \frac{\zetavect^{\mvect}}{(\mvect - \yvect)!} \right]    \label{eq:w-pois} \\
& \propto & \! \left\langle \sum_{y}  \zetavect^{\mvect-\yvect} \frac{\mvect!}{(\mvect - \yvect)!} \omega_y, \amat(\kappa) \Psi(\zetavect) \right\rangle ,  \nonumber
\eea
where $\langle , \rangle$ is the standard inner product in $\mathbb{R}^\nc$ and $\omega_y$ is a unit vector with the component corresponding to $y$ equal to 1 and all others 0; the set of $\omega_y$ for all complexes $y$ is a basis for $\mathbb{R}^\nc$.  The second line of Eq.~\ref{eq:w-pois} makes clear the connection between the existence of a PSSD and the DRE in the form (\ref{eq:dre-matrices}).

For a given network topology and choice of the $\kapyyp$, there is a PSSD if $\mastw \pois(\mvect; \zetavect) = 0$ for some $\zetavect$ and all $\mvect$.  Clearly, it is sufficient that the following condition hold:
\be
\text{There exists } \zetavect \text{ such that } \amat(\kappa) \Psi(\zetavect) = \mathbf{0}.  \label{eq:A-Psi-condit}
\ee
Because the left side of the inner product in Eq.~\ref{eq:w-pois} depends on $\mvect$ while the right side does not, one might guess that it is also necessary that the right side vanish to have $\mastw \pois = 0$ for all $\mvect$.  This is indeed the case.  In brief, one constructs a set of values of $\mvect$ such that the corresponding vectors on the left side of the inner product in (\ref{eq:w-pois}) span $\mathbb{R}^\nc$:  beginning with an $\mvect$ such that $\mvect!/(\mvect - \yvect)!$ vanishes for all but one complex $y$, one increases the components of $\mvect$ until the ratio is nonzero for two complexes, and so on until the space is spanned. (When the network has conserved quantities, the argument requires the introduction of a generalized chemical potential.)

The preceding paragraph gives our main result:  For a reaction network to have a PSSD, it is necessary and sufficient that (\ref{eq:A-Psi-condit}) hold.   The same equation appears in the theory of DRE's.  From Eq.~\ref{eq:dre-matrices}, it is clear that if there is a concentration vector $\cvect^*$ such that $\amat(k) \Psi(\cvect^*) = \mathbf{0}$, this $\cvect^*$ is a fixed point of the dynamical system $\dot{\cvect} = \fvect(\cvect)$.  
It turns out that one can say considerably more~\cite{feinberg}.  Specifically, if there exists such a $\cvect^*$, it is the unique fixed point (for given values of any conserved quantities), and it comes equipped with a Lyapunov function $\mathcal{L}(\cvect) = \sum_{j=1}^{\ns} c_{\cha_j} \ln(c_{\cha_j}/c_{\cha_j}^*) - (c_{\cha_j} - c_{\cha_j}^*)$.  This strongly resembles the free energy of a mixture of ideal gases with concentrations $c_{\cha_j}$.  Thus, strikingly, those reaction networks that have an equilibrium-like Poisson SSD when stochastic effects are taken into account are precisely those that minimize a free-energy-like function in the DRE limit.

The theory of Feinberg and coworkers~\cite{feinberg} also gives an easy way to identify many networks that satisfy (\ref{eq:A-Psi-condit}).  Every network can be assigned a topological index, the deficiency $\delta \equiv \nc - l - s$.   Here, $s$ is the dimension of the \textit{stoichiometric subspace} S of $\mathbb{R}^{\ns}$ spanned by all vectors $\yvect - \yvect'$ with $\kapyyp \neq 0$, and $l$ is the number of \textit{linkage classes}.  Two complexes $y$ and $y'$ are in the same linkage class if there is a chain of reactions $y \rightarrow y_1 \rightarrow y_2 \rightarrow \ldots \rightarrow y'$ linking either $y$ to $y'$ or $y'$ to $y$.  A network is said to be \textit{weakly reversible} if whenever there is such a chain from $y$ to $y'$ there is also one from $y'$ to $y$ (Fig.~\ref{fig1}).

References~\cite{feinberg} show that (\ref{eq:A-Psi-condit}) holds for any weakly reversible network with $\delta = 0$; such networks thus all have a PSSD.  What sort of networks are these?  Fig.~\ref{fig1}a--e gives some examples.  Since $s \leq \ns$ is a measure of the number of independent species in the network, we can interpret the requirement $\delta = 0$ as a sort of sparseness condition:  In a deficiency zero network, the average molecule cannot react with too many different sets of partners---that is, be in too many different complexes.  It is this sparseness that gives these networks their equilibrium-like character.  For an equilibrium network, we can define the standard free energy difference between two species by summing the log ratio of forward to backward rates along a reaction path that converts one species into the other.  One can think of $\ln(\zeta_{\cha_j})$ in the PSSD for a deficiency zero network as being constructed in a roughly similar manner.  The difference is that in the one case detailed balance guarantees that a consistent $\zetavect$ can be found no matter which path we take between species.  In the other it is instead the sparseness enforced by $\delta = 0$ that ensures that no inconsistencies arise. 

We next turn to some extensions of our central result describing the set of networks with a PSSD.  The ability to satisfy a fluctuation-dissipation theorem (FDT) is often taken as a hallmark of equilibrium-like behavior~\cite{durian,liu-ohern}.  It is less well-appreciated that any system governed by a master equation obeys a similar linear response relation~\cite{graham}.  In general, if the SSD $\pss(\mvect; \mu)$ depends on a parameter $\mu$, the observable conjugate to $\mu$ is $O_{\mu}(\mvect) \equiv (k_\text{B} T) \, \partial \ln[ \pss(\mvect;\mu)]/\partial \mu$.  That is, for any other observable $O'(\mvect)$, the correlation function $\langle O'(t) O_{\mu}(0) \rangle$ determines the linear response of $O'$ to a time-dependent $\mu(t)$, just as in the FDT.  A driven system can thus be said to obey the FDT if the conjugate parameter-observable pairs ($\mu$,$O_\mu$) are those expected from equilibrium reasoning.  For a system with a PSSD, this is easily shown.  We focus here on the simple case where $\mu$ plays the role of a chemical potential, but 
the generalization to other parameters is straightforward.

To model contact with a particle reservoir, we add to any network the reactions $\phi \rightleftarrows \ymu$ with rates $\kappa_{+}$ to create and $\kappa_{-}$ to destroy particles, where $\ymu$ is a complex with only 1 species.  A large enough reservoir will remain in equilibrium even when brought into contact with a driven reaction network; its chemical potential then turns out to be $\mu = \ln(\kappa_{+}/\kappa_{-})$ plus constants that depend only on the system volume and $\ymu$'s composition.    One can readily show that if the original network with stoichiometric subspace S had a PSSD, the new system in contact with the reservoir also has a PSSD provided $\yvect_{\mu} \not\in \text{S}$.  The parameters $\zetavect_{\mu}$ characterizing the new PSSD are related to the parameters $\zetavect$ of the old PSSD by $\zetavect_{\mu} = \zetavect e^{\mathbf{u}}$.  Here the $j^{\text{th}}$ component of $\zetavect e^{\mathbf{u}}$ is $\zeta_j e^{u_j}$, and $ \mathbf{u} \propto \mathbf{\ymu}^{\perp}(\mu + \text{constants})$, with $\mathbf{\ymu}^{\perp}$ the projection of $\mathbf{\ymu}$ onto the orthogonal complement $\text{S}^{\perp}$ of S. Substituting $\zetavect_{\mu}$ into the expression (\ref{eq:pssd}) defining a PSSD, we see that the observable conjugate to the chemical potential $\mu$ is $\mathbf{\ymu}^{\perp} \cdot \mvect$, just as in an equilibrium system.  The driven system thus appears to satisfy the FDT.

\begin{figure}

\includegraphics[width=6cm]{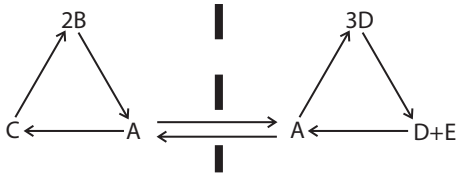}

\caption{ \label{fig2} Two reaction networks separated by a semi-permeable membrane through which they exchange A molecules.  At steady state, A has the same (generalized) chemical potential on both sides.}

\end{figure}

A related phenomenon occurs if two reaction vessels are brought into contact and allowed to exchange molecules of species A (Fig.~\ref{fig2}).  Define the (generalized) chemical potential $\mu$ of the molecules in a given vessel by adding the reactions $\phi \rightleftarrows \cha$ as in the previous paragraph and choosing $e^{\mu} \propto \kappa_{+}/\kappa_{-}$ so that the average molecule number of each species remains unchanged.   One can show that the A molecules in the two vessels have the same chemical potential at steady state, again in perfect analogy to equilibrium thermodynamics.

So far, we have dealt only with reactions in well-stirred systems.  It turns out that our results remain true even without stirring.  Specifically, consider a system in which the same reaction network is replicated at each site on a lattice, and a molecule $\cha_j$ can hop to any neighboring site with a rate $\lambda_j$.  If the reaction network at each site in isolation has a PSSD, the same is true of the coupled system on the lattice, whose SSD takes the form $\pss(\mvect_1,\mvect_2,\mvect_3,\ldots) = (1/Z) \zetavect^{\sum_{\alpha} \mvect_{\alpha}}/(\prod_{\alpha} \mvect_{\alpha}!)$, where $\mvect_{\alpha}$ gives the number of each species on lattice site $\alpha$.  To verify this result, simply observe that the terms proportional to each $\lambda_j$ in $\mastw \pss$ separately sum to zero.

Finally, we note that, as in the deterministic case~\cite{sontag}, our results extend to certain models with non-mass-action kinetics:  If the contribution $m_{\cha_j}!/(m_{\cha_j} - y_{\cha_j})!$ of the species $\cha_j$ to the reaction rate of the complex $y$ is everywhere replaced by $\prod_{i = m_{\cha_j} - y_{\cha_j} + 1}^{m_{\cha_j}} w_{\cha_j}(i)$, for some function $w_{\cha_j}$, then the PSSD generalizes to a product over species of the form $\prod_{\cha_j} \pi_{\cha_j}(m_{\cha_j})$.  Such models arise as approximate descriptions of networks with enzymatically catalyzed reactions; the product form of the SSD implies that noise in one species does not feed through the network to other species downstream~\cite{levine-hwa}.  Our results thus extend~\cite{levine-hwa} to reactions involving more than one reactant.  


In sum, we have given the first characterization of the set of chemical reaction networks whose stationary state is a Poisson distribution, as it would be for a system in equilibrium.  It turns out that a large class of networks have this property, including all those that are weakly reversible and are sufficiently sparse, in the sense that they have $\delta = 0$.  Whether $\delta$ vanishes depends only on which reactant species can be turned into which product species; a network with the appropriate topology will have a PSSD for any values of its rate constants.   This contrasts with the equilibrium case, where varying a rate generally drives a network out of equilibrium without any change in its topology.  

The significance of these results is severalfold.  In light of the tremendous interest in noise in biochemical systems, there is a need to strengthen the theoretical foundations of the subject.  In particular, there are very few exact results for non-trivial networks~\cite{wolynes}; in most cases, the only techniques available are simulations or expansions about the DRE.   This paper substantially increases our arsenal of solvable models.  It appears that few functional biological networks have $\delta = 0$; indeed, it is possible that biology avoids such architectures in part because their PSSD enforces a variance in molecule number of order the mean.  Nonetheless, one can imagine many uses for our results.  The most obvious is as a starting point for a perturbative treatment of networks near those with a PSSD; this would represent a set of approximations independent of expansions in noise strength, and thus would complement such standard approaches.

As important as any biological applications is the new perspective our work gives on the physics of driven systems.  There has long been interest in equilibrium-like descriptions of non-equilibrium phenomena, and chemical networks with a PSSD represent a novel set of examples along these lines.  In more extensively-studied granular systems, a common intuition is that a thermal description should apply when one energy scale dominates and can be used to define an effective temperature.  One can think  of this energy scale as arising from a balance between energy input and dissipation.  Chemical reaction networks suggest a generalization of this scenario.  A thermal analogy holds when the number of interactions available to a given species, and thus the number of channels to inject or dissipate energy, is limited.  This defines an energy scale for each species; differences between these scales are accommodated by our freedom to choose $\ln(\zeta_{\cha_j})$, which plays the role of an internal free energy per molecule, for each $\cha_j$.  It is tempting to speculate that the equilibrium analogy could be extended to other driven systems with a similar flexibility to define different energy scales for different components. 

I am grateful to Eduardo Sontag for helpful discussions.  This work was funded in part by NSF grant PHY05-51164 to the KITP and by FOM/NWO.

\end{document}